# Aboriginal Astronomical Traditions from Ooldea, South Australia, Part 2: Animals in the Ooldean Sky

Trevor M. Leaman

School of Humanities & Languages, University of New South Wales, NSW, 2052, Australia
Email: t.leaman@unsw.edu.au

Duane W. Hamacher

Monash Indigenous Centre, Monash University, Clayton, VIC, 3800, Australia
Email: duane.hamacher@monash.edu

and

Mark T. Carter

DesertLife Bird Guiding and Wildlife Surveys, Alice Springs, NT, 0870, Australia
Email: mark@desertlife.com.au

**Abstract**

Australian Indigenous astronomical traditions hint at a relationship between animals in the skyworld and the behaviour patterns of their terrestrial counterparts. In our continued study of Aboriginal astronomical traditions from the Great Victoria Desert, South Australia, we investigate the relationship between animal behaviour and stellar positions. We develop a methodology to test the hypothesis that the behaviour of these animals is predicted by the positions of their celestial counterparts at particular times of the day. Of the twelve animals identified in the Ooldean sky, the nine stellar (i.e. non-planet or non-galactic) associations were analysed and each demonstrated a close connection between animal behaviour and stellar positions. We suggest that this may be a recurring theme in Aboriginal astronomical traditions, requiring further development of the methodology.

> *"Most of the totemic ancestral beings are represented in the sky by stars and planets. Although they leave their material bodies on earth metamorphosed into stone, their spirits are the stars."*
>
> – Ronald and Catherine Berndt (1943).

## 1 INTRODUCTION

The study of the astronomical knowledge and traditions of Indigenous Australians has revealed a wealth of traditional knowledge regarding the night sky. Calendars and food economics are closely integrated with astronomical traditions (Fredrick, 2008; Johnson, 1998; Sharpe, 1993) often involving animals and their behavioural habits. Oral traditions describing animals in the skyworld are common across Australia (Stanbridge, 1861) and the world (e.g. Kelly and Milone, 2011: 499; Urton, 1981). These animals may be represented by constellations, asterisms, individual stars, star clusters, planets, nebulae, or other celestial objects. In Australia, celestial animals are commonly linked to behavioural patterns of their terrestrial counterparts, such as mating, birthing, or brooding their young (Cairns and Harney, 2003; Johnson, 1998; Stanbridge, 1861). These traditions serve, in part, as a guide for noting the time of year to access particular food sources.

This paper is a continuation of the study of Aboriginal astronomical knowledge in the Great Victoria Desert near Ooldea, South Australia (see Leaman and Hamacher, 2014 for Part 1). Most of this information comes from the work of amateur anthropologist Daisy Bates (Bates, 1904-1935; 1921a,b; 1924a,b; 1933; 1938) and professional anthropologists Ronald and Catherine Berndt (Berndt, 1941, Berndt and Berndt, 1943; 1945; 1974; 1977). The primary information used in this paper comes from Daisy Bates in a story she recorded about the constellation and star cluster known by Western astronomers as Orion and the Pleiades (Bates, 1921b; 1933). Through her work, Bates described many animals in the Ooldean sky, but gave no details about a relationship between the animal and its celestial counterpart. In most cases, the stories simply stated the type of animal that each celestial object represented and their major or minor role in the narrative.

In this paper, we test the hypothesis that each animal in the Ooldean sky is associated with a celestial object that is used to predict the breeding habits of these animals, such as mating, birthing, incubating eggs, brooding, and fledging young[1]. Specifically, we investigate whether the breeding habits of these animals are predicted by the heliacal or acronychal rising or setting, or meridional transit, of their respective celestial counterpart.

## 2 ANIMALS IN THE ABORIGINAL SKYWORLD

According to the cosmogony and cosmography of Aboriginal Australians, the realm of the skyworld has topography similar to and every bit as real as the terrestrial landscape below (Clarke, 2007/2008; 2015b). This realm is inhabited by plants, animals and ancestral beings, each represented by celestial bodies (Clarke, 2014; 2015a,b; Leaman & Hamacher, 2014), or other prominent features of the night sky, such as the prominent dark bands of the Milky Way (e.g. see Fuller et al, 2014a,b).

In his foundational work on Aboriginal astronomy in western Victoria, William E. Stanbridge (1857; 1861) recorded some of the astronomical traditions handed to him by the Boorong, a clan of the Wergaia language group living near Lake Tyrell. His papers include several animals that relate to stars. For instance, the star Vega (α Lyrae), called *Neilloan* in the Wergaia language, was linked to the Mallee fowl (*Leipoa ocellata*), a chicken-sized ground-dwelling megapode that builds its nest-mounds when Vega rises at dusk (acronychal rising). When Vega is high in the sky at dusk (dusk meridian crossing), the birds are laying their clutches of eggs, and when Vega sets at dusk (heliacal setting), the chicks begin hatching. Similarly, the star Arcturus (α Boötis), called *Marpeankurrk* in the Wergaia language, is related to the larvae of the wood-ant, which are plentiful for only a couple of months of the year – August and September - the same time Arcturus is visible in the evening sky. In his Master of Arts thesis and subsequent research, historian John Morieson (1996; 1999) built upon these associations to construct a detailed picture of the Boorong night sky, noting the complex calendars that are related to seasonal animal behaviour.

Recent studies of Aboriginal astronomical knowledge (e.g. Cairns and Harney, 2003; Fredrick, 2008; Fuller et al., 2014b; Hamacher, 2012) show a definitive link between animal behaviour and the positions of their celestial counterparts in the sky, particularly at dusk and dawn. The appearance of the celestial emu in the evening sky, traced out by the dark spaces in the Milky Way from Crux to Sagittarius, informs Aboriginal people of when emus are laying their eggs, which is an important food source (Fuller et al., 2014a; Norris and Hamacher, 2009).

Examples like these are commonplace in Aboriginal traditions, yet it seems most of the traditions collected by early anthropologists do not provide much information of this sort. This paper explores animals in the Ooldean sky for connections to their terrestrial counterparts in terms of annual breeding behaviour patterns to better understand the nature of Aboriginal astronomical knowledge.

## 3 ANIMALS IN THE OOLDEAN SKY

Ooldea is on the traditional lands of the Kokatha peoples (Gara, 1989). The presence of a permanent water source at Ooldea Soak made it an important drought refuge and meeting place for trade and ceremony for these people and many surrounding Aboriginal language groups (Bates, 1938; Gara, 1989; Tindale, 1974: 6). At the time of Daisy Bates' visit to Ooldea, activities associated with the Trans-Australian Railway (Bates, 1938; Brockwell et al, 1989; Colley et al, 1989) was causing major disruption to traditional lifestyles of these peoples, an important one being the establishment of more permanent camps, with diverse peoples of the region living in close proximity to each other. This may explain the blending of vocabularies from different language groups in the word lists of Daisy Bates (e.g. see Bates, 1918), and in the skylore she recorded in the Ooldea region (Bates, 1921a,b, 1924a,b, 1933)[2]. Here, rather than attempting to disentangle Bates' records, we have adopted the term 'Ooldean sky' to describe the linguistically-blended skylore of the region at the time of its recording by Bates.

The records from Bates (1904-1935) note several animals in the Ooldean sky (Table 1): the bush turkey, black cockatoo, dingo, emu, grey kangaroo, owlet nightjar, crow, redback spider, red kangaroo, thorny devil lizard, and the wedge-tailed eagle. All animals are related to stars, star clusters, asterisms or the dark band of Milky Way (see also Leaman and Hamacher, 2014). The exceptions are the red and grey kangaroos, which are related to the "morning" and "evening" stars, respectively. Bates identifies these as the planets Jupiter and Venus, respectively, but it is not clear if Bates is misrepresenting Jupiter as the morning star (as Venus is typically called both the morning and evening star) or if Jupiter was prominent in the morning when Bates recorded the relevant astronomical traditions. There is no time stamp for the dates she recorded these traditions, but they are in her notes from Ooldea, where she lived and worked from 1919 to 1935. During that 16 year period, both Venus and Jupiter would have been bright in the early morning sky on multiple occasions. The black cockatoo is related to both Antares and Mars. Planets are not suitable for denoting *annual* seasonal change on Earth, but may be used in referencing longer climatic cycles (e.g. droughts, floods, El Niño Southern Oscillation driven events, etc.), so Venus, Mars, and Jupiter will have no regular connection with the annual seasonal behavioural cycles of any animal.

A list of the celestial objects, their location in the sky, their spectral type (which indicates colour), and their brightness (visual magnitude, $m_V$) are given in Table 2. All of the individual stars listed are first magnitude (among the brightest 22 stars in the night sky). We exclude *Babba*, the dingo father, as we do not know the exact star that represents Babba in the sky. Bates describes it as being the 'horn of the bull' (Bates, 1933). Leaman and Hamacher (2014) argue that it could be either the stars β or ζ Tauri, but this is uncertain, and Aldebaran (α Tauri) is already ascribed to a major character in the narrative.

*Table 1: A list of animals in the Ooldean sky and their celestial counterpart in alphabetical order, taken from "The Orion Story" recorded by Daisy Bates (Leaman & Hamacher, 2014).*

| **Aboriginal Name** | **Animal** | **Species** | **Type** | **Object** |
|---|---|---|---|---|
| *Gibbera* | Australian Bustard | *Ardeotis australis* | Bird | Vega |
| *Joorr-Joorr* | Owlet Nightjar | *Aegotheles cristatus* | Bird | Canopus |
| *Kalia* | Emu | *Dromaius novaehollandiae* | Bird | Coalsack |
| *Kangga Ngoonji* | Crow Mother | *Corvus spp.* | Bird | Altair |
| *Kara* | Redback Spider | *Latrodectus hasseltii* | Arachnid | Arcturus |
| *Kogolongo* | Black Cockatoo | *Calyptorhynchus banksii* | Bird | Mars |
| *Kulbir* | Grey Kangaroo | *Macropus fuliginosus* | Mammal | Venus |
| *Maalu* | Red Kangaroo | *Macropus rufus* | Mammal | Jupiter |
| *Ngurunya (?)* | Dingo | *Canis lupus dingo* | Mammal | Achernar |
| *Nyumbu* | Crow (chicks) | *Corvus spp.* | Bird | Delphinus |
| *Waljajinna* | Wedge-Tailed Eagle | *Aquila audax* | Bird | Crux |
| *Warrooboordina* | Black Cockatoo | *Calyptorhynchus banksii* | Bird | Antares |
| *Yugarilya* | Thorny Devil Lizard | *Moloch horridus* | Reptile | Pleiades |

*Table 2: Data for each of the celestial objects mentioned in Table 1. Data includes common name, Bayer designation, coordinates (right ascension and declination in J2000). Spectral types, and visual magnitudes ($m_V$) are taken from the SIMBAD star database (simbad.u-strasbg.fr). For planets the range of $m_V$ are the minimum to maximum values. The value given for the coordinates of Crux is 35 Crucis, which lies near the centre of the constellation. The $m_V$ quoted are for α Crucis and δ Crucis, the brightest and faintest of the four major stars in Crux. Planet colours are approximate.*

| Name | Designation/ Type | RA | DEC | SpecType | Colour | $m_V$ |
|---|---|---|---|---|---|---|
| Achernar | α Eridani | 01h 37m 43s | −57° 14’ 12” | B6V | Blue-White | 0.46 |
| Altair | α Aquilae | 19h 50m 47s | +08° 52’ 06” | A7V | White | 0.76 |
| Antares | α Scorpii | 16h 29m 24s | −26° 25’ 55” | M0.5Iab | Orange | 0.91 |
| Arcturus | α Bootis | 14h 15m 40s | +19° 10’ 57” | K0III | Orange | −0.05 |
| Canopus | α Carinae | 06h 23m 57s | −52° 41’ 44” | A9II | Yellow-White | −0.74 |
| Coalsack | Dark Nebula | 12h 31m 19s | −63° 44’ 36” | – | - | – |
| Crux | (Constellation) | 12h 31m 40s | −59° 25’ 26” | – | - | 0.81(α), 2.78(δ) |
| Delphinus | (Constellation) | 20h 39m 38s | +15° 54’ 43” | – | - | > 3.00 |
| Jupiter | (Planet) | – | – | – | *Red-White* | −1.61 to −2.94 |
| Mars | (Planet) | – | – | – | *Red* | 1.84 to −2.91 |
| Pleiades | (Open Star Cluster) | 03h 47m 00s | +24° 07’ 00” | A0, B6, B7 | Blue-White | 1.60 |
| Vega | α Lyrae | 18h 36m 56s | +38° 47’ 01” | A0V | White | 0.03 |
| Venus | (Planet) | – | – | – | *White-Yellow* | −3.8 to −4.9 |

## 4 TERRESTRIAL BEHAVIOUR OF OOLDEAN SKY ANIMALS

The behaviour patterns and lifecycles of the animals represented in the Ooldean sky are discussed in the following sections. These will set the foundation on which we test the connections between animal behaviour and stellar positions.

### 4.1 AUSTRALIAN BUSTARD (VEGA)

The star Vega (α Lyrae) is *Gibbera*, the bush turkey (Figure 2a). In the deserts of Central Australia, the term bush turkey, or brush turkey, is colloquially used by Aboriginal people to refer to the Australian bustard (*Ardeotis australis*). This is not to be confused with the Australian brush (bush) turkey (*Alectura lathami*), which inhabits more temperate and wet tropical areas. The bustard is an important food source for Aboriginal people (Ziembicki, 2009).

The breeding cycle of the Australian bustard varies across Australia and appears to be closely linked to weather and seasonal patterns, especially rainfall frequency (Ziembicki, 2009; Ziembicki and Woinarski, 2007, Figure 1). In arid areas, such as at Ooldea, populations are transient and migratory, with numbers fluctuating in response to habitat and food availability in wet/dry years (Ziembicki, 2009). In good years, breeding at Ooldea generally occurs between May and August, with a slight peak in June (Ziembicki, 2009). Chicks emerge after an incubation period of 23 days (Beruldsen, 2003).

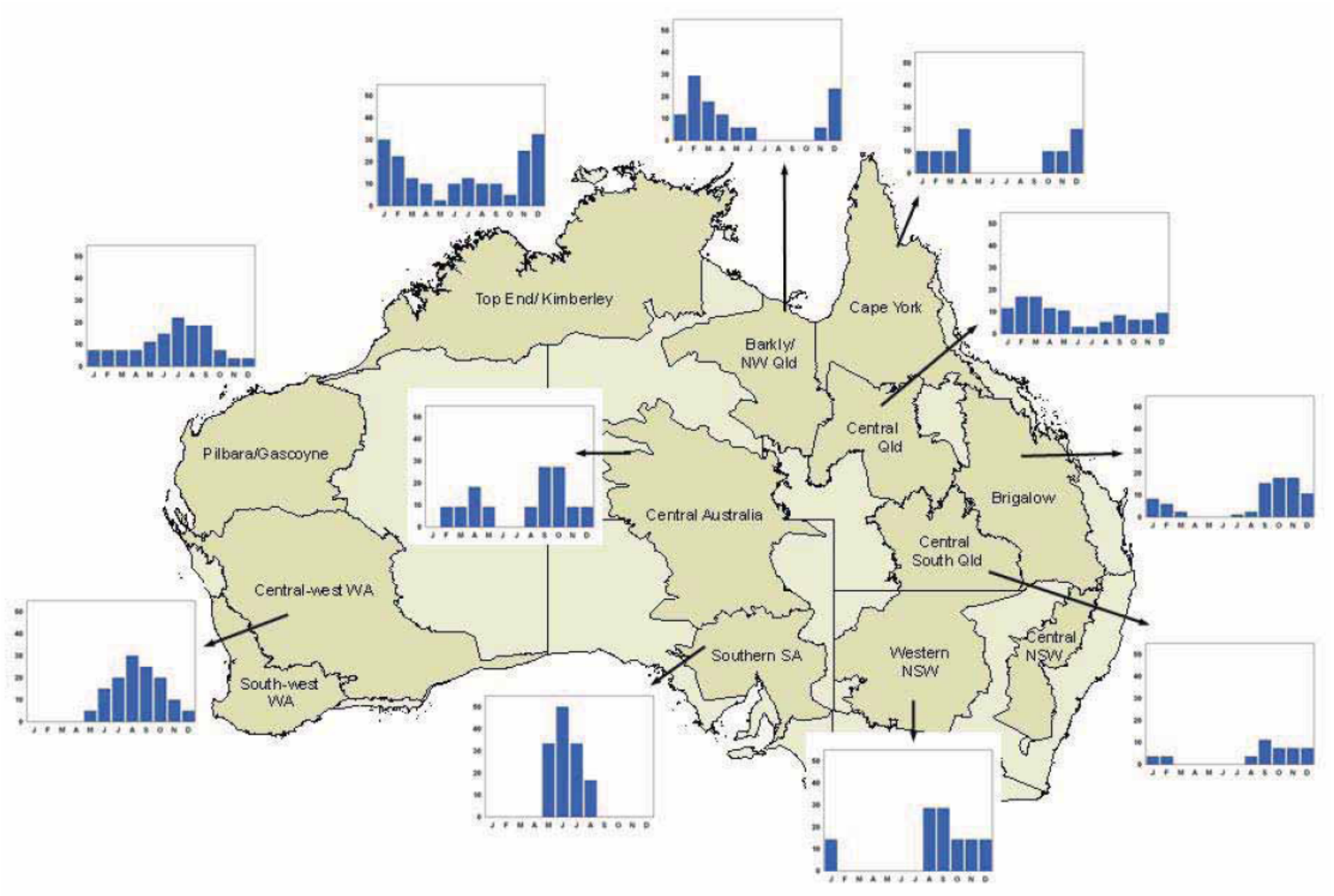

*Figure 1: Regional variation in breeding cycle peaks in the Australian bustard in response to seasonal variations in rainfall patterns over a year, in terms of mm precipitation (y-axis) by month (x-axis) (from Ziembicki, 2009: 49). Ooldea lies at the extreme western edge of the area labeled 'Southern SA'.*

## 4.2 CROW MOTHER (ALTAIR) AND HER CHICKS (DELPHINUS)

The star Altair ($\alpha$ Aquilae) is *Kangga Ngoonji*, the mother crow, and the stars of the constellation Delphinus are *Nyumbu,* her chicks. In the Pitjantjatjara language, Kangaa (or Kaanka) is the name for the Torresian Crow (*Corvus orru*) (Reid et al, 1993). However, since this species is not normally found south of the Birksgate Ranges, to the northeast of the Great Victoria Desert, it is unlikely that this is the species referred to in Bates' story. A similar looking species, the Little Crow (*Corvus bennetti*) does inhabit the area around Ooldea, and is easily mistaken for the Torresian crow. Although the Pitjantjatjara name, *Wangarangara*, is normally specific to this species, there is a chance that the name *Kangaa Ngoonji* was used for it around the time Bates was recording her story, especially if her informants were less particular with distinguishing between the two. To add more confusion, another similar looking species, the Australian Raven (*Corvus coronoides*) (Figure 2b) can also be found at Ooldea. It inhabits a thin stretch of land across the Nullarbor and in the desert regions of South Australia (Beruldsen, 2003).

Regardless of which species is referred to in Bates' account, the breeding season is similar for all three, being from July to September, and the incubation period for the clutch is roughly 20 days. The chicks are fledged within 45 days, but the mother continues to feed them for up to four months (Beruldsen, 2003).

### 4.3 EMU (COALSACK NEBULA)

The Coalsack Nebula is the head of *Kalia* the emu (*Dromaius novaehollandiae*) (Figure 2c). The Coalsack is a dark absorption nebula that borders Crux, Centaurus, and Musca, and resembles the profile of an emu's head, replete with beak. Emus form into breeding pairs in December and January and begin mating in late April to early May and continue through to June (Eastman, 1969). Emus lay one egg every day or two and incubation does not commence until all eggs are laid. As clutch sizes vary between five to 20 eggs, it can take up to three weeks for the whole clutch to be laid. Incubation of the eggs lasts for 56 days (Eastman, 1969; Reid et al., 1993). Male emus rear the chicks for up to seven months. Interestingly, wedge-tailed eagles are the prime predators of emu chicks (Reid et al., 1993), which may be another reason that the constellation of the Wedge-tailed eagle has been placed in Crux, adjacent to the celestial Emu, its natural prey in both celestial and terrestrial worlds.

The association of the Coalsack with the head of the emu is found across Australia (e.g. Cairns and Harney, 2003; Fuller et al., 2014a; Stanbridge, 1861; Wellard, 1983). Although only the Coalsack is specified as "the emu", in many Aboriginal traditions it is the head of the emu, with the eye represented by the star BZ Crucis ($m_V$ = 5.3) (Hamacher, 2012: 71). A profile view of the emu is traced out along the Milky Way, from the Coalsack to the centre of the galaxy in Scorpius, Ophiuchus, and Sagittarius, where the galactic bulge outlines the body of the emu. Similarly, Bates (1904-1935: No. 25/308, p. 13) claims that the "long dark patch in the Milky Way" is the "emu father" in the traditions of another desert community. Since the emu is traced out by a large part of the sky, the criterion used for single celestial objects is not applicable, hence the Coalsack's absence in Tables 3 and 4. Similarly, we exclude it from further analysis here. However, we reiterate that across Australia, the rising of the celestial emu at dusk coincides with the breeding and egg-laying season of the emu (e.g. Fuller et al, 2014a).

### 4.4 BLACK COCKATOO (ANTARES)

The red-giant star Antares (α Scorpii) is *Warrooboordina*, the red-tailed black cockatoo (*Calyptorhynchus banksii*) (Figure 2d). The red-tailed black cockatoo is the only species of the genus *Calyptorhynchus* found in the Central Desert. Although its distribution does not extend to the Ooldea region, probably due to the lack of mature River Red Gums along major watercourses that the species requires for breeding, they do occur through the Musgrave Ranges, near the northern extent of the Anangu-Pitjantjatjara-Yankunytjatjara (APY) lands ~480 km to the north of Ooldea. The red colour of the cockatoo's tail feathers provides a clue as to why it is connected to the bright red star Antares. These birds breed from March to September (Forshaw, 2002) and incubate their clutch for around 30 days (Kurucz, 2000). Breeding for the inland subspecies *Calyptorhynchus banksii samueli* also begins in March but has a peak in July (Higgins, 1999; Storr, 1977). Fledging takes a median of 87 days after hatching and chicks are fed by both parents for a further three to four months after leaving the nest (Higgins, 1999).

## 4.5 OWLET NIGHTJAR (CANOPUS)

The star Canopus (α Carinae) is Joorr-Joorr, the Australian owlet nightjar (*Aegotheles cristatus*) (Figure 2e), found throughout the Australian Outback wherever tree hollows or rock crevices are present in which the birds can breed. The bird's name is onomatopoeic, mimicking the sound of one of its repertoire of nocturnal calls. This and other calls commonly used by this species have been described as sounding like human 'laughs' or 'chuckles' (Higgins, 1999: 1042-1043).

This bird breeds mainly from October to January with an incubation period of 18 to 29 days, depending on temperature (Brigham and Geiser, 1997). The chicks from the first clutch of eggs hatch by late October and are fledged about a month afterwards. The birds also go through a short period of hibernation (torpor) from May to September (Bringham, et al. 2000).

## 4.6 DINGO (ACHERNAR)

The star Achernar (α Eridani) is *Ngurunya*, the mother dingo (*Canis lupus dingo*) (Figure 2f). Across most parts of Australia, the dingo breeding season generally begins in March with gestation lasting between 61 and 69 days (Corbett, 1995), with the first litters being whelped from May to July (Catling *et al.,* 1992). Although there is a distinct breeding peak in the Central Desert in March, dingoes have been observed with pups all year round (Purcell, 2010: 43), suggesting that breeding cycles are also likely influenced by seasonal availability of food resources due to weather and climate cycles.

## 4.7 THORNY-DEVIL LIZARD (PLEIADES)

The Pleiades (M45 open star cluster) are *Yugarilya*, the seven Mingari sisters. As Nyeeruna (Orion) attempts to seduce the sisters (Leaman and Hamacher 2014), they become frightened and transform into the thorny devil lizard (*Moloch horridus*) (Figure 2g). The thorny devil is their totem animal and plays a central role in the narrative.

The lizard, also called the "thorny dragon", is small (20 cm in length) with conical spines and camouflaged skin (Browne-Cooper et al., 2007). Females lay a clutch of ten eggs in a burrow some 30 cm deep between the months of September and December (Pianka, 1997). The eggs incubate for three to four months (*ibid*), after which the lizards crawl out of their subterranean burrows.

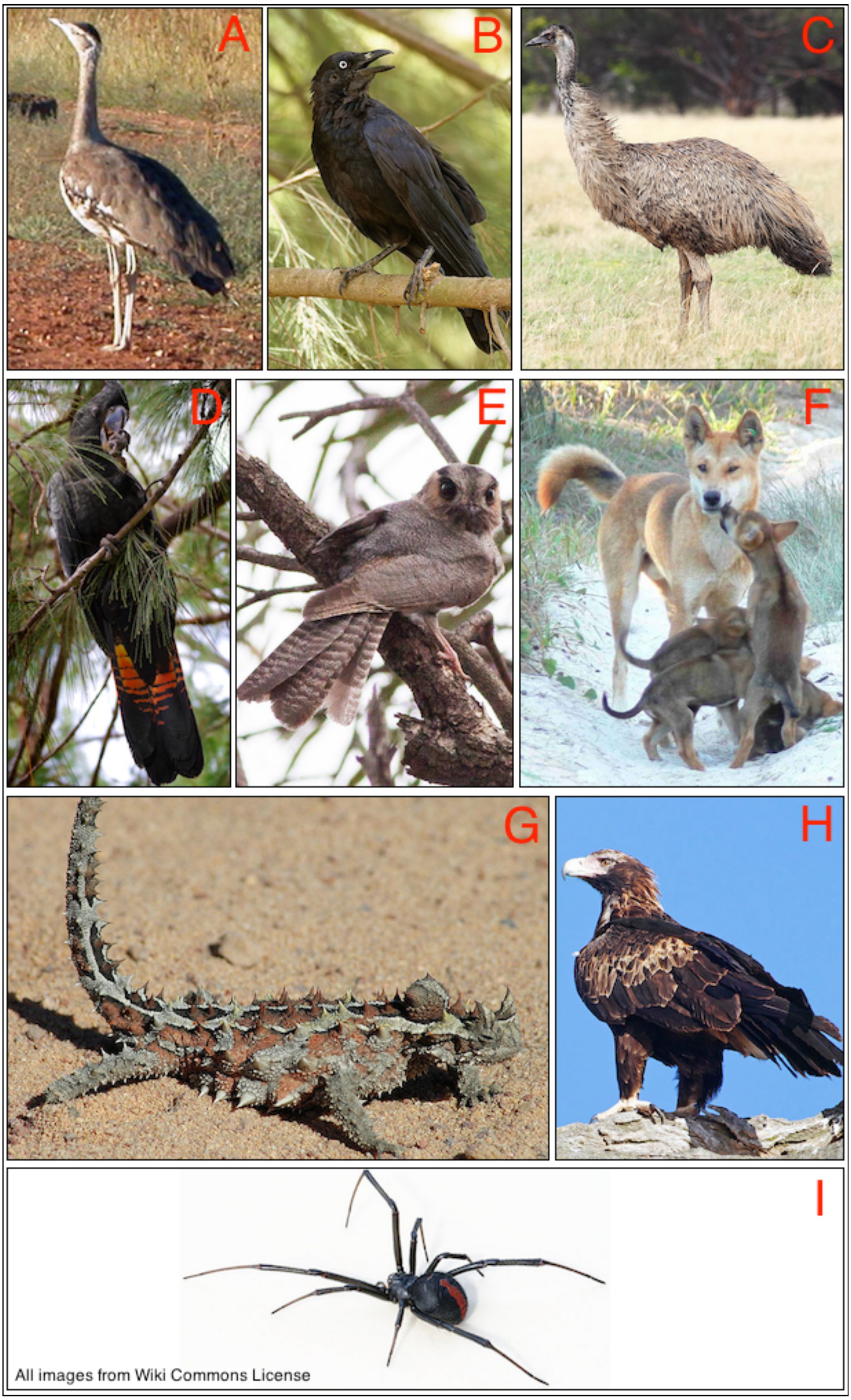

*Figure 2: Animals in the Ooldean sky (images reproduced from Wikimedia Commons license): A) Australian bustard (Glen Fergus), B) Australian raven (J.J. Harrison), C) emu (Benjamint444), D) red-tailed black cockatoo, E) owlet nightjar (Ron Knight), F) dingo male with pups (Partner Hund), G) thorny devil lizard (thorny dragon) (KeresH.), H) wedge-tailed eagle (J.J. Harrison), and I) redback spider (Toby Hudson).*

## 4.8 WEDGE-TAILED EAGLE (CRUX)

The stars of the Southern Cross constellation (Crux) are seen as the footprint of *Waljajinna*, the eagle-hawk, or wedge-tailed eagle (*Aquila audax*) (Figure 2h). The four brightest stars of Crux resemble the footprint of the wedge-tailed eagle – an association found in the astronomical traditions of Arrernte and Luritja communities in the Central Desert (e.g. Maegraith, 1932: 20; Mountford, 1976: 41).

The wedge-tailed eagle is one of the largest birds of prey in the world and is commonly found across the Ooldea region. The breeding season begins in March and April and runs to September, but a majority of the eagles lay their eggs in July (Olsen, 1995, 2005). Eggs are incubated for 45 days.

Normally two eggs are laid, but it is very common for only one of these to survive to fledge the nest. The weaker chick is usually out-competed for food and sometimes killed by the stronger sibling, a phenomenon known as "Cainism" after the Old Testament siblicide story of Cain and Able. This is a common action among eagles in the *Aquila* genus worldwide (Olsen, 1995; Simmons, 1998). The weaker chick is thought to usually be dead within 20 days of hatching. Fledging rates are variable, between 75 to 95 days, dependent on factors such as food availability and nest disturbance (Marchant and Higgins, 1993).

## 4.9 REDBACK SPIDER (ARCTURUS)

The red-orange star Arcturus (α Boötis) is *Kara*, the redback spider (*Latrodectus hasseltii*) (Figure 2i). Although identified in some of Bates' records as the blue star Rigel (β Orionis), this was probably a transcription error in her notes (for justification of this conclusion see Leaman and Hamacher, 2014).

Redback spiders, which are less common in desert regions, mate year round. However, the breeding rate increases when the temperature is warmer, peaking in summer (Forster, 1995). Studies near Perth, WA showed that redback spider sexual activity begins to peak in late November (Andrade, 2003). The spiderlings may emerge from the sacs as early as 11 days after being laid, but their emergence is also temperature-dependent, and cooler temperatures may prolong the time before they emerge.

# 5 METHODOLOGY

How can we test for connections between a particular star or asterism given animal associations and the behaviour of its terrestrial counterpart when these connections are not explicitly stated by Aboriginal people or included in the written record? We develop a methodology that considers two primary variables: the time and duration of each particular animal behaviour and the time and position of the associated star's appearance in the sky. Animal behaviour, particularly breeding, may be dependent on weather, temperature, and other variables. As a consequence, the timing of a particular animal behaviour can range on the order of weeks. Conversely, the rise or set of stars at particular times from a given location in a given year can be calculated to the exact day.

The development of an appropriate methodology should make best use of the information available in the most rigorous manner possible. A similar study relating stellar positions to seasonal changes was published by Hamacher (2015). Building upon this methodology, we attempt to predict the reasons why each animal is linked to its celestial counterpart.

Traditionally, archaeoastronomers use precise definitions for heliacal and acronychal rising and setting (e.g. Aveni, 1980, 2001; Bruin, 1979; Robinson, 2009; Schaefer, 1987, 1997). According to these definitions, *heliacal rise* (HR) occurs when an object is first visible above the eastern horizon just before the sun rises. *Heliacal set* (HS) occurs when an object is last visible above the western horizon after sunset. Similarly, *acronychal rise* (AR) occurs when an object is first visible above the eastern horizon just after sunset, and *acronychal set* (AS) occurs when an object is last visible above the western horizon just before sunrise. A star's highest point in the sky occurs when a star crosses (transits) the meridian (an imaginary line across the sky connecting due North and South). For this study, we define meridional transit at sunrise (Mdawn) or sunset (Mdusk) as the point when the star crosses the meridian at its last and first visibility, respectively. All six aspects are illustrated for clarity in Figure 3.

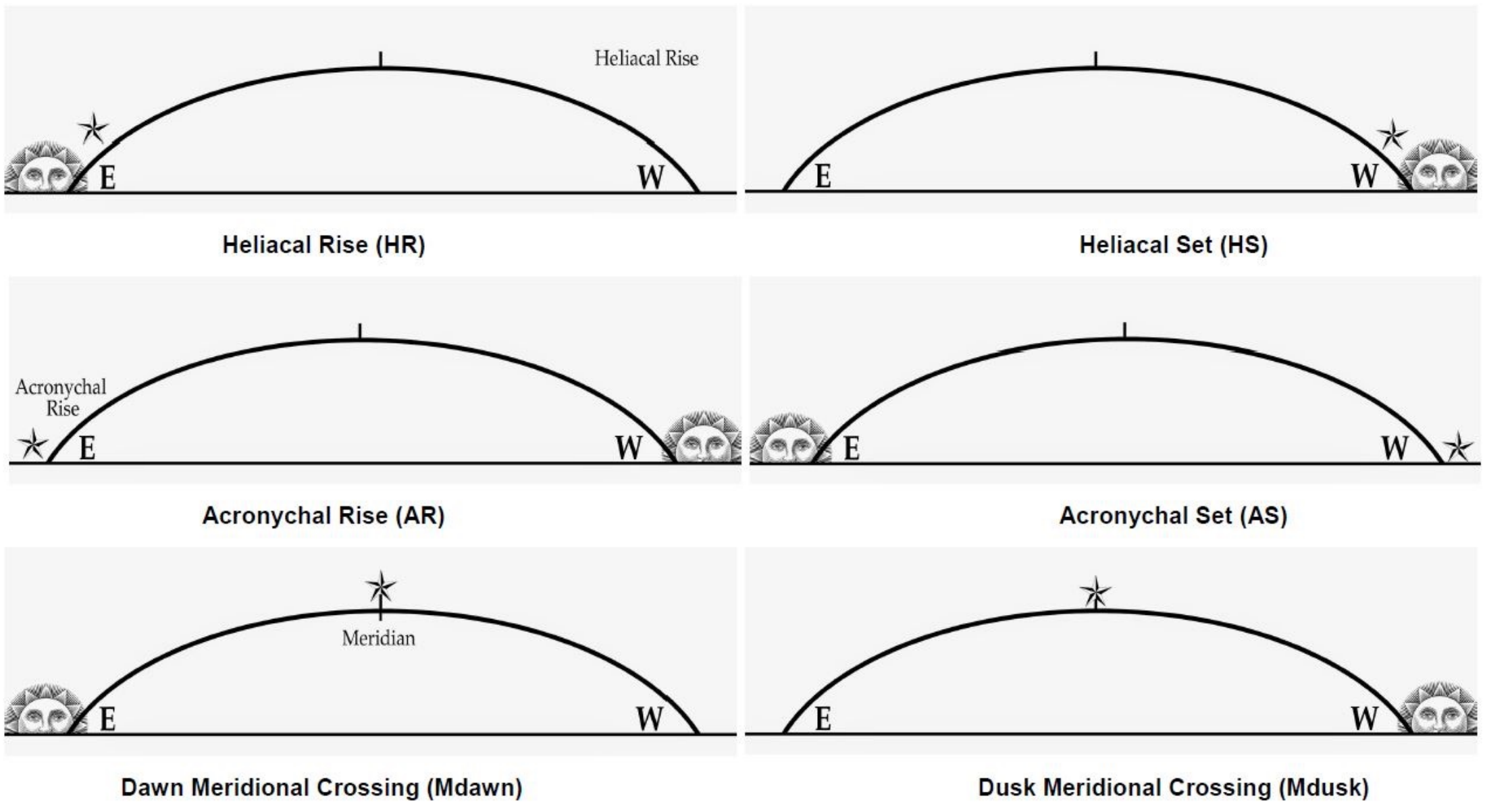

*Figure 3: A diagrammatical representation of the six aspects described in the text. Image courtesy of Freedom Cole. URL: svatantranatha.blogspot.com.au/2014/09/heliacal-cycle.html*

The heliacal, acronychal, or meridional visibility of a celestial object depends on several factors, including its brightness relative to the background glow of the sun (Aveni, 2001: 112), refraction of the starlight and sunlight through the Earth's atmosphere, atmospheric conditions, extinction, visual acuity of the observer, visibility of the horizon (e.g. vegetation and local topography), and the elevation and altitude of the horizon from the observer (see Hamacher, 2012: 70-71; Hamacher, 2015). The interested reader is directed to Aveni (1980, 2001), Schaefer (1986, 1987, 1997, 2000) and Robinson (2009) for a more detailed treatise on observing conditions and visibility as applied to cultural astronomy research.

For a best case scenario, which assumes no moon glow, ideal observing conditions (dry, no cloud cover, still air, etc.), and an acute observer (20/10 Snellen vision ratio), 1st magnitude stars become visible at heliacal rise/set at a minimum altitude of 5° (the extinction angle, see Schaefer, 2000) when the sun has an altitude of −10°. For 2nd magnitude stars, the sun's altitude needs to be a minimum of −14°, and −16° for 3rd magnitude stars (Aveni 1980: 110). For example, to an observer at Lake Tyrrell in western Victoria in the year 2000, the heliacal rising of the Pleiades occurred on 18 June. On this day, the Pleiades rose at 05:36 AM (azimuth at 0°), emerged from Earth's obscuring atmosphere to become visible at 06:07 AM (Pleiades at +5° azimuth, Sun at −16° azimuth) and were visible for a further 50 minutes (Pleiades at +13°, Sun at – 7°) before the sun drowned out the light of the star cluster.

Here, we test for connections between the rising/setting or meridian transit times of stars having animal associations (as recorded by Bates and the Berndts), and the behaviours and lifecycles of their terrestrial animal counterparts. We utilise the Stellarium positional astronomy software package[3] for calculations of stellar positions throughout the year using the coordinates of Ooldea, South Australia (30° 27′ 0″ S, 131° 50′ 0″ E, elevation of 80 m; see Tables 3 and 4). As we have no time stamp of the dates Bates recorded these stories, azimuths and altitudes are calculated for the year 1919, being the first year of Bates' arrival at Ooldea and the earliest possible date she could have collected these traditions.

The effects of precession, nutation, and stellar proper motion, or the range of observing locations across the Ooldea region, are negligible for this study and will not greatly affect the values cited here. The rise and set times are based on *apparent* azimuths and altitude, not *geometric* azimuths and altitudes (the former values take into account atmospheric refraction).

The heliacal/acronychal rise and set times can be calculated to the day, as would be necessary for an archaeoastronomical study on the precise alignments of ancient monuments. But that is not necessary for the purpose of this study. There is a significant window of time between the exact rising/setting times of these stars and the breeding behaviour of the animals (which can range from days to months). Studies (as cited above) show that Aboriginal astronomical knowledge relating stars to animals was not intended to be precise "to-the-day", but rather to simply inform the people of the seasonal habits of these animals. The values cited in this paper are meant to provide an approximate measurement of the time, date, and location in the sky these objects rise and set, and their connection to animal breeding cycles.

Taking all factors into consideration, the criteria used here for estimating the date of each respective astronomical event is given as follows:

- For the 1st magnitude stars Achernar, Altair, Antares, Arcturus, Canopus, and Vega:

- o Heliacal or acronychal rise/set occurs when each star has a minimum altitude of 5° and the sun has a maximum altitude of −10°.
- o Meridional transit occurs when each star crosses the meridian when the sun has a maximum altitude of −10°.

- The constellation Crux, which comprises stars ranging from 1st to 3rd magnitude, will be treated as a 2nd magnitude star focused on γ Crucis (as α Crucis is circumpolar).

  Therefore:

  - o Heliacal or acronychal rise/set occurs when γ Crucis has a minimum altitude of 5° and the sun has a maximum altitude of −14°.
  - o Meridional transit occurs when γ Crucis crosses the meridian when the sun has a maximum altitude of −14°.

- Delphinus, which contains stars ranging from 3rd to 5th magnitude, and the Pleiades, will be treated as a 3rd magnitude star located at α Delphinus and η Tauri (Alcyone) respectively.

  Therefore:

  - o Heliacal or acronychal rise/set occurs when α Delphinus or η Tauri has a minimum altitude of 5° and the sun has a maximum altitude of −16°.
  - o Meridional transit occurs when α Delphinus or η Tauri crosses the meridian when the sun has a maximum altitude of −16°.

## 6 RESULTS AND ANALYSIS

The stellar aspect data presented in Tables 3 and 4 was combined with the annual lifecycle information outlined in Section 4 to form a metadata table, with the year broken down into twelve months of four weeks each (see Table A1, Appendix). From this table, the connection between the stellar aspect and the start of a particular lifecycle stage (e.g. mating/laying; birthing/hatching; whelping/fledging, etc.) was graded according to "closeness of fit", and given a colour code accordingly (see Table 5):

- Green = stellar aspect occurs within two weeks of the start of a lifecycle stage;
- Yellow = between three and four weeks;
- Red = between five and six weeks;
- Grey = more than six weeks.

Where there is an extended breeding cycle (such as in the Black Cockatoo and Wedge Tail Eagle), or the cycle is continuous throughout the year (such as in the case of the Dingo), the connection was determined from the beginning of the peak in these cycles. The following sections discuss the observed connections between stellar aspect and lifecycle stage for each of the terrestrial animals represented in the Ooldean sky.

*Table 3: Acronychal rising (AR) and setting (AS) azimuths (in degrees), dates, and times for celestial objects representing Ooldean animals. The altitude (in degrees), date, and time of when each object crosses the meridian after sunset (Mdusk) is also provided. η Tauri and α Delphinus are used to calculate times and positions for the Pleiades and Delphinus, respectively, and γ Crucis is used for Crux (as α Crucis is circumpolar). The times are calculated for when the object has an altitude of at least 5°. Values calculated using Stellarium.*

| Object | Acronychal Rise (AR) | | | Dusk Meridian Crossing (Mdusk) | | | Acronychal Set (AS) | | |
|---|---|---|---|---|---|---|---|---|---|
| | Az | Date | Time | Alt | Date | Time | Az | Date | Time |
| Achernar | 159° | 23-Aug | 19:35 | 62.9° | 18-Dec | 21:03 | 201° | 31-Dec | 5:22 |
| Altair | 76° | 28-Jul | 19:22 | 51.0° | 07-Oct | 20:01 | 283° | 06-Jul | 7:23 |
| Antares | 118° | 19-May | 19:11 | 85.8° | 23-Aug | 19:35 | 243° | 08-Jun | 7:18 |
| Arcturus | 63° | 14-May | 19:13 | 40.0° | 24-Jul | 19:20 | 296° | 14-Apr | 6:47 |
| Canopus | 151° | 02-Nov | 20:22 | 67.8° | 12-Mar | 20:19 | 209° | 20-Feb | 6:10 |
| Crux | 157° | 07-Jan | 21:33 | 63.7° | 25-Jun | 19:27 | 203° | 21-May | 6:49 |
| Delphinus | 68° | 6-Aug | 19:55 | 43.9° | 11-Oct | 20:32 | 291° | 22-Jul | 6:51 |
| Pleiades | 58° | 12-Nov | 21:02 | 35.6° | 09-Jan | 21:43 | 302° | 05-Dec | 4:39 |
| Vega | 49° | 04-Aug | 19:26 | 20.7° | 21-Sep | 19:50 | 321° | 26-May | 7:10 |

*Table 4: Heliacal rising (HR) and setting (HS) azimuths (in degrees), dates, and times for celestial objects representing Ooldean animals. The altitude (in degrees), date, and time of when each object crosses the meridian before sunrise (Mdawn) is also provided. η Tauri and α Delphinus are used to calculate times and positions for the Pleiades and Delphinus, respectively, and γ Crucis is used for Crux (as α Crucis is circumpolar). The times are calculated for when the object has an altitude of at least 5°.*

| Object | Heliacal Rise (HR) | | | Dawn Meridian Crossing (Mdawn) | | | Heliacal Set (HS) | | |
|---|---|---|---|---|---|---|---|---|---|
| | Az | Date | Time | Alt | Date | Time | Az | Date | Time |
| Achernar | 159° | 11-Mar | 6:26 | 62.9° | 15-Jul | 7:22 | 201° | 06-Jun | 19:05 |
| Altair | 76° | 16-Feb | 6:06 | 51.0° | 25-Apr | 6:42 | 283° | 11-Dec | 20:59 |
| Antares | 118° | 18-Dec | 5:15 | 85.8° | 11-Mar | 6:26 | 243° | 17-Nov | 20:36 |
| Arcturus | 63° | 14-Dec | 5:14 | 40.0° | 12-Feb | 6:01 | 296° | 26-Sep | 19:55 |
| Canopus | 151° | 23-May | 7:09 | 67.8° | 20-Oct | 5:46 | 209° | 02-Aug | 19:24 |
| Crux | 157° | 23-Aug | 6:38 | 63.7° | 26-Jan | 5:22 | 203° | 24-Oct | 20:33 |
| Delphinus | 68° | 08-Mar | 5:53 | 43.9° | 12-May | 6:35 | 291° | 10-Dec | 21:34 |
| Pleiades | 58° | 16-Jun | 6:50 | 35.6° | 01-Sep | 6:19 | 302° | 19-Apr | 20:04 |
| Vega | *38*° | 22-Feb | 6:11 | 20.7° | 8-Apr | 6:44 | 321° | 05-Nov | 20:26 |

*Table 5: Connection of Stellar aspect (AR = acronychal rise; AS = acronychal set; HR = heliacal rise; HS = heliacal set; Mdawn = dawn meridian crossing; Mdusk = dusk meridian crossing) with the annual lifecycles of terrestrial Ooldean animals. Colours denote degree of connection (see legend). *Aspect order in each row follows the sequence through the year from 1st Jan to 31st Dec.*

| Star | Animal | Lifecycle | Connection between Stellar Aspect and Lifecycle* | | | | | |
|---|---|---|---|---|---|---|---|---|
| Achernar | Dingo Mother | Mating | HR | HS | Mdawn | AR | Mdusk | AS |
| | | Birthing | HR | HS | Mdawn | AR | Mdusk | AS |
| | | Whelping | HR | HS | Mdawn | AR | Mdusk | AS |
| Altair | Crow Mother | Mating/Laying | HR | Mdawn | AS | AR | Mdusk | HS |
| | | Hatching | HR | Mdawn | AS | AR | Mdusk | HS |
| | | Fledging | HR | Mdawn | AS | AR | Mdusk | HS |
| Antares | Black Cockatoo | Mating/Laying | Mdawn | AR | AS | Mdusk | HS | HR |
| | | Hatching | Mdawn | AR | AS | Mdusk | HS | HR |
| | | Fledging | Mdawn | AR | AS | Mdusk | HS | HR |
| Arcturus | Red Back Spider | Mating/Laying | Mdawn | AS | AR | Mdusk | HS | HR |
| | | Hatching | Mdawn | AS | AR | Mdusk | HS | HR |
| Canopus | Owlet Nightjar | Mating/Laying | AS | Mdusk | HR | HS | Mdawn | AR |
| | | Hatching | AS | Mdusk | HR | HS | Mdawn | AR |
| | | Fledging | AS | Mdusk | HR | HS | Mdawn | AR |
| | | Torpor | AS | Mdusk | HR | HS | Mdawn | AR |
| Crux | Wedge-Tailed Eagle | Mating/Laying | AR | Mdawn | AS | Mdusk | HR | HS |
| | | Hatching | AR | Mdawn | AS | Mdusk | HR | HS |
| | | Fledging | AR | Mdawn | AS | Mdusk | HR | HS |
| Delphinus | Crow Chicks | Laying | HR | Mdawn | AS | AR | Mdusk | HS |
| | | Hatching | HR | Mdawn | AS | AR | Mdusk | HS |
| | | Fledging | HR | Mdawn | AS | AR | Mdusk | HS |
| Pleiades | Thorny Devil | Mating/Laying | Mdusk | HS | HR | Mdawn | AR | AS |
| | | Hatching | Mdusk | HS | HR | Mdawn | AR | AS |
| Vega | Bustard | Mating/Laying | HR | Mdawn | AS | AR | Mdusk | HS |
| | | Hatching | HR | Mdawn | AS | AR | Mdusk | HS |
| | | Fledging | HR | Mdawn | AS | AR | Mdusk | HS |

| Connection occurring: | |
|---|---|
| (green) | Within 1 to 2 weeks |
| (yellow) | Between 3 to 4 weeks |
| (red) | Between 5 - 6 weeks |
| (grey) | More than 6 weeks |

## 6.1 AUSTRALIAN BUSTARD (VEGA)

The acronychal rising (AR) of Vega occurs in early August and sets at sunset (HS) by mid-November, crossing the meridian (reaching its maximum altitude) in the northern sky at dusk in late September. Therefore, the star is prominent in the evening skies during the entire bustard mating season for most regions of Australia (Figure 2), with peaks occurring close to the dusk meridian transit (Mdusk). A similar relationship is given regarding the Mallee fowl and the star Vega in Wergaia traditions of western Victoria (Stanbridge, 1861).

According to the population survey conducted in 2007-2009, the breeding cycle in the region surrounding Ooldea can occur earlier, from May to July (Ziembicki, 2009). In this case, mating/laying/hatching corresponds to the acronychal setting (AS) of Vega, and fledging occurs during the star's acronychal rise (AR). As the bustard population is transient and highly dependent on rainfall, it is likely that the actual breeding patterns surrounding Ooldea are highly variable, and wetter years may be more typical of other regions. In the absence of other reliable data, we rely on that supplied by Ziembicki (2009) for this study.

## 6.2 CROW MOTHER (ALTAIR) AND HER CHICKS (DELPHINUS)

The acronychal rising (AR) of Altair occurs in late July, when the crows begin laying clutches of eggs. By early August the eggs are hatching and the brighter group of stars in Delphinus (the chicks) rise at sunset (AR). By October, when most chicks are fledged, Altair crosses the meridian in the sky after sunset (Mdusk). In December, as the last of the chicks start leaving the nest, both Altair (mother crow) and Delphinus (celestial chicks) set at dusk (HS). It is also worth noting that Altair and Delphinus first set in the western sky at dawn (AS) during the very start of the breeding season in July, before reappearing in the eastern sky at dusk (AR). This gives two possible (and sequential) connections to signify the start of the breeding season.

## 6.3 BLACK COCKATOO (ANTARES)

The dawn meridian crossing (Mdawn) of Antares coincides with the start of the breeding cycle, and the acronychal rising (AR) in early May coincides with the first clutches of eggs hatching. The peak in the breeding cycle occurs soon after the acronychal setting (AS) in mid-June. By mid-August, towards the end of the breeding season and the start of fledging, Antares crosses the meridian and is nearly at zenith after sunset (Mdusk).

In the traditions recorded by Bates, the red-tailed black cockatoo was also called *Kogolongo*, represented by the planet Mars. The connection of both objects to this bird is almost certainly due to the red feathers in the bird's tail. Their relationship to each other is most likely due to the fact that the ecliptic passes through Scorpius and Mars sometimes comes to within 6° of Antares[4]. This occurred ten times in the evening sky between 1919 and 1935 (September 1920: angular separation = 2.7°, March 1922: 5.4°, July 1922: 2.4°, February 1924: 4.9°, January 1926: 4.7°, December 1927: 4.4°, December 1929: 4.2°, November 1931: 3.9°, October 1933: 3.8°, and September 1935: 3.1°).

It is unclear if this provides us with a time-stamp of when Bates recorded the story, but we can reasonably assume that the red colour of Mars and Antares and their occasional close approach are the reasons for both objects being associated with the red-tailed black cockatoo.

### 6.4 OWLET NIGHTJAR (CANOPUS)

The dawn meridian crossing (Mdawn) and acronychal rising (AR) of Canopus occurs in mid and late October, respectively, and is high in the southern sky at dawn, coinciding with the start of the breeding season. In the "Orion Story" (Leaman and Hamacher, 2014) Joorr-Joorr observes Nyeeruna's attempts to seduce and impress the seven Mingari sisters, represented by the Pleiades, and laughs at Nyeeruna's humiliation at the hands of the eldest sister Kambugudha, represented by the Hyades. By early December, when most chicks are fledging and begin using the 'churring' adult call (Higgins, 1999), Orion is rising at dusk. A combination of fledglings calling and a seasonal spike in the bird's vocal activity during the warmer nights of summer, (Schodde and Mason, 1980) leading to increased vocalization overall. This may explain why Joorr-Joorr laughs at Nyeeruna in the "Orion Story" (Leaman and Hamacher, 2014).

Another interesting celestial correspondence is the heliacal rise (HR) of Canopus just as the Owlet Nightjar begins a period of winter torpor. This lifecycle concludes over the onset of spring, coinciding with the heliacal set (HS) of Canopus.

### 6.5 DINGO (ACHERNAR AND THE PLEIADES)

The peak of the breeding season in March corresponds to the heliacal rising (HR) of Achernar to the southeast. By June, when the dingoes begin whelping pups, Achernar sets at dusk (HS) and the Pleiades rise at dawn (HR). Dingoes are related to the Mingari women of the Pleiades who kept a "tribe of dingoes" with them to keep the men away (Bates, 1933). In Aboriginal cultures of the Central Desert, the heliacal rising of the Pleiades signalled the start of winter and the time to start harvesting dingo puppies (e.g. Clarke, 2007: 51; Mountford, 1956, 1958).

## 6.6 THORNY-DEVIL LIZARD (PLEIADES)

The start of egg-laying coincides with the dawn meridian crossing (Mdawn) of the Pleiades. The emergence of the first clutch of lizards from their nests in early December is heralded by the acronychal rising (AR) of the Pleiades just two weeks prior to this, followed by the acronychal setting (AS) closer to the actual hatching time, with the last hatchings occurring just prior to the heliacal set (HS) in April.

Their most important predator is the bustard (Pianka and Pianka, 1970). Interestingly, this predator-prey relationship may be seen in the interaction between their celestial counterparts: As Vega (Bustard) disappears from view in the northwest sky, the Pleiades (thorny devil) 'safely' emerges soon afterwards in the northeast. This scenario is similar to the eternal pursuit of Orion and Scorpius from Greek mythology.

## 6.7 WEDGE-TAILED EAGLE (CRUX)

Crux is at its highest altitude in the sky at dusk (Mdusk), with $\alpha$ and $\gamma$ Crucis crossing the meridian at close to the same time) in late June, coinciding with the start of the peak in breeding and laying. This also coincides with the hatchings of the first clutches of eggs laid in late March, with the siblicide process usually completed and the remaining chick well on the way to fledging. The peak in breeding then carries over to a peak in hatching in mid- to late-August, coinciding with the heliacal rise (HR), and a peak in fledging in mid- to late-November, just after heliacal set (HS).

## 6.8 REDBACK SPIDER (ARCTURUS)

The peak in the breeding cycle of Redback spiders in late November results in the majority of spiderlings emerging from the egg sacs in early to mid-December, closely corresponding to the heliacal rising (HR) of Arcturus.

## 7 DISCUSSION AND CONCLUDING REMARKS

We focus on the associations between the celestial animals of the Ooldea night sky and the lifecycle of their terrestrial counterparts. Excluding planetary (e.g. Venus, Mars, Jupiter) and galactic (e.g. 'celestial emu') associations, we demonstrate that each of the remaining nine celestial animals represented appear to inform Aboriginal people about the lifecycle (e.g. mating/breeding, laying/birthing, fledging/whelping, etc.) of their terrestrial counterparts. More specifically, we found that the 26 lifecycles studied correlated with stellar aspects in the proportions as follows (Figure 4):

- acronychal rising (9),
- acronychal setting (8),
- dusk meridian crossing (4),
- heliacal rising (5),
- heliacal setting (3) and
- dawn meridian crossing (5)

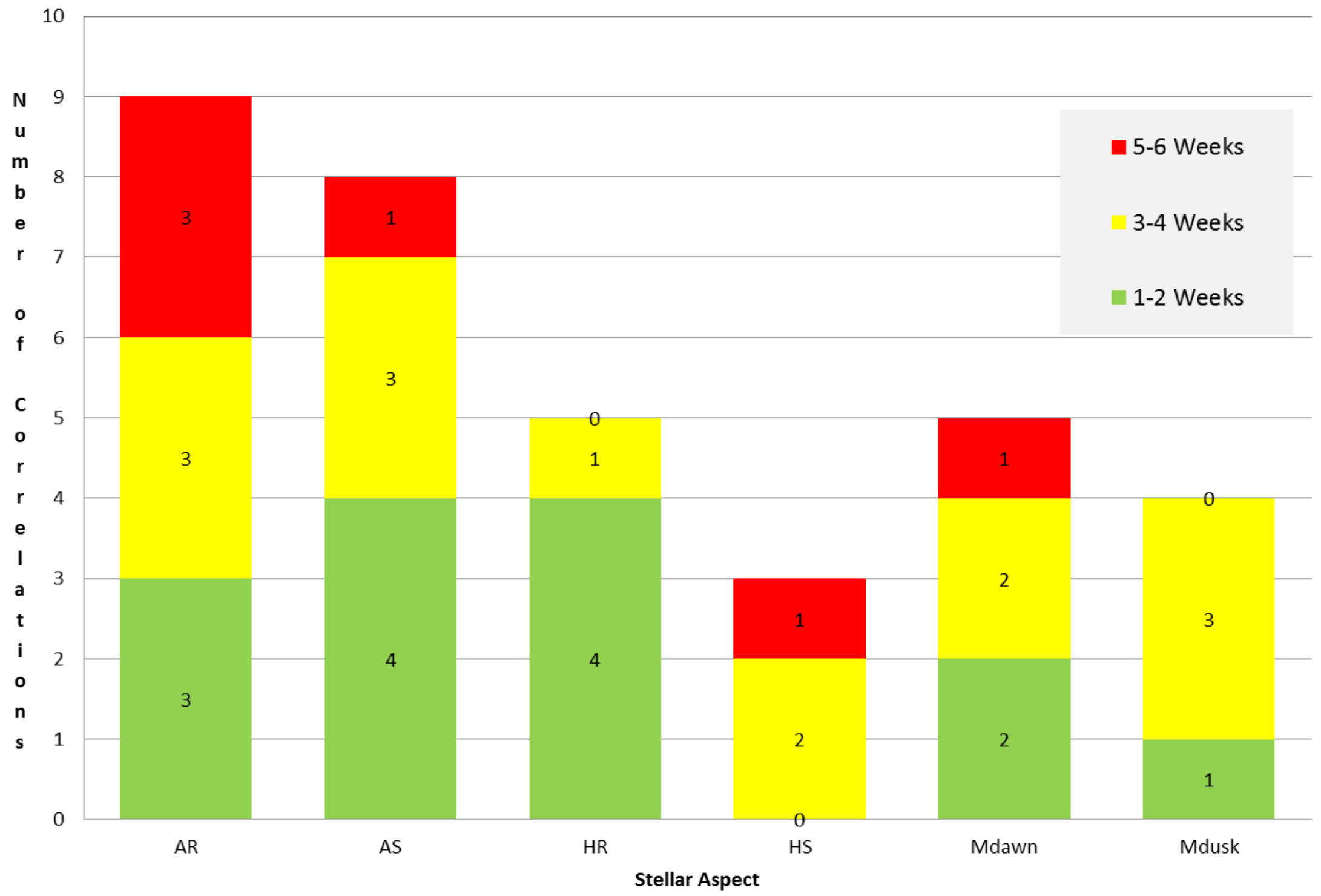

*Figure 4: Summary table of total numbers of connections between stellar aspect of the Ooldean animals and the lifecycles of their terrestrial counterparts. (AR = acronychal rise; AS = acronychal set; HR = heliacal rise; HS = Heliacal set; Mdawn = dawn meridional crossing; Mdusk = dusk meridional crossing).*

Of these 34 correlations, 14 (41%) occur within two weeks of one of these six stellar aspects, and 28 (82%) occur within four weeks. Similarly, there appeared to be a stronger bias towards lifecycles occurring within four weeks of an acronychal aspect (AR, AS: 46%), than for meridional (Mdawn, Mdusk: 29%) or heliacal (HR, HS: 25%) aspects.

This study suggests that Aboriginal people from Ooldea deliberately selected certain prominent stars and asterisms to match the breeding cycles of the terrestrial animals they represent. Further studies will determine whether this is (a) a unique feature of Ooldean astronomy, (b) restricted to a few language groups, or (c) a more common thread running through Aboriginal astronomical traditions across Australia. A larger study is planned to address these questions.

Implicit in this study is a method of determining the antiquity of these stellar-terrestrial connections. Firstly, as lifecycles of some animals are dependent on temperature (e.g. Red Back Spider, Thorny Devil) and/or rainfall (e.g. Bustard, Dingo) (see Section 4), then Australian climate variability constrains these connections to the post-glacial period of the last ~9,000 years, over which Australia started to experience temperature and rainfall patterns similar to current trends (e.g. see Reeves et al, 2013). Secondly, the effects of precession further constrain this to the last ~2,000 years. Any earlier than this and precessional effects would cause out-of-sync mismatches between stellar aspects and terrestrial lifecycles such that the selected star would be an unreliable calendrical indicator, and another star or asterism would need to be assigned (Hamacher, 2012).

The results of this study are not definitive, as the meaning of the star/animal relationship is known to the original Ooldean Aboriginal custodians of the traditions and not the authors of this paper. The effects of colonisation on Indigenous traditions and knowledge systems have been severely negative, and some traditions have been damaged, fragmented, or lost altogether. The development of methodologies used to reconstruct astronomical traditions is simply a step forward in helping Indigenous communities reclaim their traditions.

## 8 ACKNOWLEDGEMENTS

We dedicate this paper to the current descendants and ancestors of all Aboriginal Australians, for they are true astronomy pioneers. We also acknowledge the knowledge custodians who shared their skylore with Bates, the Berndts, and other researchers. They are the primary source and owners of this knowledge.

We thank Professor Wayne Orchiston and the two anonymous referees whose invaluable feedback assisted us in improving this paper. We also acknowledge the contributions from students in *ATSI 3006: The Astronomy of Indigenous Australians*, an undergraduate course developed and taught by Hamacher at the University of New South Wales in 2014 and 2015. In particular, we thank Eddie Ip and John Willauer in the 2015 class for making connections between stars and animal behaviour initially overlooked by the authors.

Leaman acknowledges support from the Australian Postgraduate Award (APA) and Hamacher acknowledges support from the Australian Research Council (DE140101600). Hamacher began this research as a staff member at UNSW, but completed the research and writing as a staff member at Monash. Leaman was a teaching assistant for ATSI 3006 and aided the students in their assessment related to this paper.

## 9 NOTES

1. Bates' Ooldean taxonomy is consistent with information in Read et al (1993). Vertebrate species names are drawn from Census of South Australian Vertebrates, 4th (current) Edition (https://data.environment.sa.gov.au/Content/Publications/Census-of-SA-Vertebrates-2009.pdf). Invertebrate taxonomy (redback spider) was drawn from Gray, M., *Laterodectus hastletti*. *Species Bank*. Australian Biological Resources Study. Canberra. (www.environment.gov.au/biodiversity/abrs/online-resources/species-bank/index.html). It should be noted that while ethnotaxonomies and zoological Linnaean taxonomies do not always agree, in the case of the species discussed in this paper, Ooldean and scientific taxonomies do overlap, with the possible exception of the Corvids (Crows/Ravens). An enlightening exploration of the complexities in reconciling Anangu and zoological taxonomy can be found in Read et al (1993: 86-88).
2. For instance, Bates' word for Emu, *Kalia,* is specific to the central and western desert regions (Pitjantjatjara language), whereas the word for Wedge-Tailed

Eagle, *Waldja(jinna)*, is from the Eyre Peninsula (Wirangu language). Many of the bird names listed in Bates' Ooldean Sky can be found in Condon (1955a,b) and Sullivan (1928), along with their language group origins.

3. Stellarium: www.stellarium.org
4. In Greek mythology, Antares was the god of war – the counterpart to the Roman god of war, Mars. Antares means "anti-Ares" or "rival of Ares", as Antares and Mars are bright red and sometimes come close together in the sky, fighting for dominance.

## 10 REFERENCES

Andrade, M.C.B., 2003. Risky mate search and male self-sacrifice in redback spiders. *Behavioural Ecology,* 14(4), 531-538.

Aveni, A.F., 1980. *Sky Watchers of Ancient Mexico*. Austin, University of Texas Press.

Aveni, A. F., 2001. *Skywatchers: a revised and updated version of Skywatchers of ancient Mexico*. Austin, University of Texas Press.

Bates, D.M., 1904-1935. *Papers of Daisy Bates*. Box 13, Section VII: Myths and Legends, Manuscript 365. Canberra, National Library of Australia.

Bates, D.M., 1918. Aborigines of the west coast of South Australia. Vocabularies and ethnographical notes. *Transactions & Proceedings of the Royal Society of South Australia*, 42, 152-167.

Bates, D.M., 1921a. The Great Plain's Edge, *The Australasian*, 17 August 1921, p. 418.

Bates, D.M., 1921b. Aborigines and Orion. *The Australasian*, 1 October 1921, p. 671.

Bates, D. M., 1924a. Aboriginal stellar myths. *The Australasian,* 26 July 1924, p. 226.

Bates, D.M., 1924b. Aboriginal astronomy. *The Sydney Morning Herald*, 22 November 1924, p. 13.

Bates, D.M., 1933. "Abo" Astronomy: The Constellation Orion. *The Sydney Morning Herald*, 9 September 1933, p. 9.

Bates, D.M., 1938. *The Passing of the Aborigines, A Lifetime spent among the Natives of Australia*. London, John Murray.

Berndt, R.M., 1941. Tribal Migrations and Myths Centring on Ooldea, South Australia. *Oceania*, 12(1), 1-20.

Berndt, R.M., and Berndt, C.H., 1943. A Preliminary Report of Field Work in the Ooldea Region, Western South Australia. *Oceania*, 14(1), 30-66.

Berndt, R.M., and Berndt, C.H., 1945. A Preliminary Report of Field Work in the Ooldea Region, Western South Australia. *Oceania*, 15(3), 239-275.

Berndt, R.M., and Berndt, C.H., 1974. *The First Australians*, 3rd Edition. Sydney, Ure Smith.

Berndt, R.M., and Berndt, C.H., 1977. *The World of the First Australians*. Sydney, Landsdowne Press.

Beruldsen, G.R., 2003. *Australian Birds: Their Nests and Eggs*. Brisbane, Gordon R. Beruldsen.

Brigham, R.M., and Geiser, F., 1997. Breeding biology of Australian owlet-nightjars *Aegotheles cristatus* in eucalypt woodland. *Emu*, 97, 316–321.

Brigham, R.M., Kortner, G., Maddocks, T.A., and Geiser, F., 2000. Seasonal use of torpor by free-ranging Australian Owlet Nightjars (*Aegotheles cristatus*). *Physiological and Biochemical Zoology*, 73(5), 613-620.

Brockwell, S., Gara, T., Colley, S. and Cane, S., 1989. The history and archaeology of Ooldea Soak and Mission. *Australian Archaeology*, 28, 55-77.

Browne-Cooper, R.; Bush, B.; Maryan, B;, and Robinson, D., 2007. *Reptiles and Frogs in the Bush: Southwestern Australia*. Perth, University of Western Australia Press.

Bruin, F., 1979. The heliacal setting of the stars and planets: II. *Proceedings of the Royal Netherlands Academy of Arts and Sciences*, 82(4), 397-410.

Cairns, H.C., and Harney, B.Y., 2003. *Dark Sparklers - Yidumduma's Aboriginal Astronomy*. Merimbula, NSW, H.C. Cairns.

Catling, P.C., Corbett, L.K., and Newsome, A.E., 1992. Reproduction in captive and wild dingoes (*Canis familiaris dingo*) in temperate and arid environments of Australia. *Wildlife Research*, 19, 195–205.

Clarke, P.A., 2007/2008. An Overview of Australian Aboriginal Ethnoastronomy. *Archaeoastronomy*, 21, 39-58.

Clarke, P.A., 2014. The Aboriginal Australian cosmic landscape, Part 1: The ethnobotany of the skyworld. *Journal of Astronomical History and Heritage*, 17(3), 307–325.

Clarke, P.A., 2015a. The Aboriginal Australian cosmic landscape, Part 2: Plant connections with the skyworld. *Journal of Astronomical History and Heritage*, 18(1), 23-37.

Clarke, P.A., 2015b. *Australian Aboriginal astronomy and cosmology*. In C.L.N. Ruggles (ed.) *Handbook of Archaeoastronomy and Ethnoastronomy*. New York, Springer. Pp. 2223-2230.

Colley, S., Brockwell, S., Gara, T. and Cane, S., 1989. The archaeology of Daisy Bates' campsite at Ooldea, South Australia. *Australian Archaeology*, 28, 78-91.

Condon, H.T., 1955. Aboriginal bird names – South Australia, part one. *South Australian Ornithologist*, July 1955, 74-88.

Condon, H.T., 1955. Aboriginal bird names – South Australia, part two. *South Australian Ornithologist*, October 1955, 91-98.

Corbett, L.K., 1995. *The dingo in Australia and Asia*. Sydney, University of New South Wales Press.

Eastman, M., 1969. *The life of the emu*. Sydney, Angus and Robertson.

Forshaw, J.M., 2002. *Australian Parrots*, 3rd Edition. Sydney, Rigby.

Forster, L.M., 1995. The behavioural ecology of Latrodectus hasselti (Thorell), the Australian redback spider (Araneae: Theridiidae): a review. *Records of the Western Australia Museum (Supplement)*, 52, 13–24.

Fredrick, S., 2008. *The Sky of Knowledge - A Study of the Ethnoastronomy of the Aboriginal People of Australia*. MPhil Thesis (unpublished). Department of Archaeology and Ancient History, University of Leicester, Leicester, UK.

Fuller, R.S., Anderson, M.G.,, Norris, R.P., and Trudgett, M.., 2014a. The Emu Sky Knowledge of the Kamilaroi and Euahlayi Peoples. *Journal of Astronomical History and Heritage*, 17(2), 171-179.

Fuller, R.S., Norris, R.P., and Trudgett, M., 2014b. The Astronomy of the Kamilaroi People and their Neighbours. *Australian Aboriginal Studies*, 2014(2), 3-27.

Gara, T., 1989. The Aborigines of the Great Victoria Desert: the ethnographic observations of the explorer Richard Maurice. *Journal of the Anthropological Society of South Australia*, 27(5), 15-47.

Hamacher, D.W., 2012. *On the Astronomical Knowledge and Traditions of Aboriginal Australians*. PhD Thesis (by publication), Department of Indigenous Studies, Macquarie University, Sydney, NSW, Australia.

Higgins, P.J. (ed), 1999. *Handbook of Australian, New Zealand and Antarctic Birds. Volume 4: Parrots to Dollarbird.* Melbourne, Oxford University Press.

Johnson, D.D., 1998. *Night Skies of Aboriginal Australia - a Noctuary*. Oceania Monograph No. 47. Sydney, University of Sydney Press.

Kelley, D.H., and Milone, E.F., 2011. *Exploring Ancient Skies: a survey of ancient and cultural astronomy*, 2nd Edition. New York, Springer.

Kurucz, N., 2000. *The nesting biology of the red-tailed black cockatoo (Calyptorhynchus banksii macrorhynchus) and its management implications in the Top End of Australia*. MSc Thesis (unpublished), School of Biological Sciences, Charles Darwin University, Darwin, NT, Australia.

Leaman, T.M., and Hamacher, D.W., 2014. Aboriginal Astronomical Traditions from Ooldea, South Australia, Part 1: Nyeeruna and the "Orion Story". *Journal of Astronomical History and Heritage*, 17(2), 180-194.

Maegraith, B., 1932. The astronomy of the Aranda and Luritja tribes. *Transactions of the Royal Society of South Australia*, 56(1), 19-26.

Marchant, S., and Higgins, P.J. (eds), 1993. *Handbook of Australian, New Zealand and Antarctic Birds. Volume 2: Raptors to Lapwings.* Melbourne, Oxford University Press.

Morieson, G.C.J., 1996. *The night sky of the Boorong: partial reconstruction of a disappeared culture in north-west Victoria*. MA Thesis (unpublished), Australian Centre, Faculty of Arts, University of Melbourne, Melbourne, VIC, Australia.

Morieson, G.C.J., 1999. The Astronomy of the Boorong. *Journal of Australian Indigenous Issues*, 2(4), 19-28.

Mountford, C.P., 1956. *Records of the American-Australian Scientific Expedition to Arnhem Land. Volume 1: Art, Myth and Symbolism*. Melbourne, Melbourne University Press.

Mountford, C.P., 1958. *The Tiwi: their art, myth, and ceremony*. Melbourne, Phoenix House.

Mountford, C.P., 1976. *The Nomads of the Australian Desert*. Adelaide, Rigby.

Norris, R.P. and Hamacher, D.W., 2009. *The Astronomy of Aboriginal Australia*. In D. Valls-Gabaud and A. Boksenberg (eds.) *The Role of Astronomy in Society and Culture*. Cambridge, University of Cambridge Press. Pp. 10-17.

Olsen, P., 1995. *Australian Birds of Prey: the Biology and Conservation of Raptors*. Sydney, University of New South Wales Press.

Olsen, P., 2005. *Wedge-Tailed Eagle*. Australian Natural History Series. Melbourne, CSIRO Publishing.

Pianka, E.R., 1997. Australia's thorny devil. *Reptiles*, 5(11), 14-23.

Pianka, E.R. and Pianka, H.D., 1970. The ecology of *Moloch horridus* (Lacertilia: Agamdae) in Western Australia. *Copeia*, 1970(1), 90-103.

Purcell, B., 2010. *Dingo*. Melbourne, CSIRO Publishing.

Reeves, J.M. et al, 2013. Climate variability over the last 35,000 years recorded in marine and terrestrial archives in the Australian region: An OZ-INTIMATE compilation. *Quaternary Science Reviews*, 74, 21-34.

Reid, J.R.W. Kerle, J.A. and Morton, S.R. (eds), 1993. *Uluru Fauna: the distribution and abundance of vertebrate fauna of Uluru (Ayers Rock - Mount Olga) National Park, NT*. Kowari, Vol. 4, Canberra, Australian National Parks and Wildlife Service.

Robinson, M., 2009. Ardua et Astra: On the Calculation of the Dates of the Rising and Setting of Stars. *Classical Philology*, 104(3), 354-375.

Schaefer, B.E., 1986. Atmospheric Extinction Effects on Stellar Alignments. *Journal for the History of Astronomy (Archaeoastronomy Supplement)*, 17, S32-S42.

Schaefer, B.E., 1987. Heliacal Rise Phenomena. *Journal for the History of Astronomy (Archaeoastronomy Supplement)*, 18, S19-S33.

Schaefer, B.E., 1997. Heliacal Rising: Definitions, Calculations, and Some Specific Cases. *Quarterly Bulletin of the Center for Archaeoastronomy*, 25, URL: http://terpconnect.umd.edu/~tlaloc/archastro/ae25.html

Schaefer, B.E., 2000. New Methods and Techniques for Historical Astronomy and Archaeoastronomy. *Archaeoastronomy*, 15, 121-136.

Schodde R., and Mason, I.J., 1980. *Nocturnal Birds of Australia*. Melbourne, Lansdowne Press.

Simmons, R., 1998. Offspring quality and the evolution of Cainism. *Ibis*, 130, 339-357.

Stanbridge, W.E., 1861. Some Particulars of the General Characteristics, Astronomy, and Mythology of the Tribes in the Central Part of Victoria, Southern Australia. *Transactions of the Ethnological Society of London*, 1, 286-304.

Storr, G.M., 1977. *Birds of the Northern Territory*. Perth, Western Australian Museum Special Publication, No. 7.

Sullivan, C., 1928. Bird notes from the West Coast. *South Australian Ornithologist*, 9(5), 164 -169.

Urton, G., 1981. Animals and Astronomy in the Quechua Universe. *Proceedings of the American Philosophical Society*, 125(2), 110-127.

Wellard, G.E.P., 1983. *Bushlore, or this and that from here and there*. Perth, Artlook Books.

Ziembicki, M., 2009. *Ecology and movements of the Australian Bustard Ardeotis australis in a dynamic landscape.* PhD Thesis (unpublished), School of Earth and Environmental Sciences, University of Adelaide, Adelaide, SA, Australia.

Ziembicki, M., 2010. *Australian Bustard.* Australian Natural History Series. Melbourne, CSIRO Publishing.

Ziembicki, M and Woinarski, J.C.Z, 2007. Monitoring continental movement patterns of the Australian Bustard Ardeotis australis through community-based surveys and remote sensing. *Pacific Conservation Biology*, 13, 128-142.

**About the Authors**

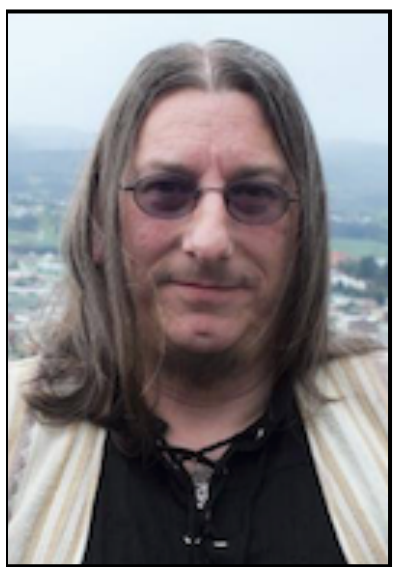

Trevor Leaman is a PhD candidate in the School of Humanities and Languages at the University of New South Wales in Sydney, Australia. He is researching the astronomical traditions of the Wiradjuri people of central NSW under the supervision of Dr Duane Hamacher and Professor Stephen Muecke. Trevor earned Bachelors degrees in biology and forest ecology, Graduate Diplomas in civil and mechanical engineering, and a Masters degree in astronomy.

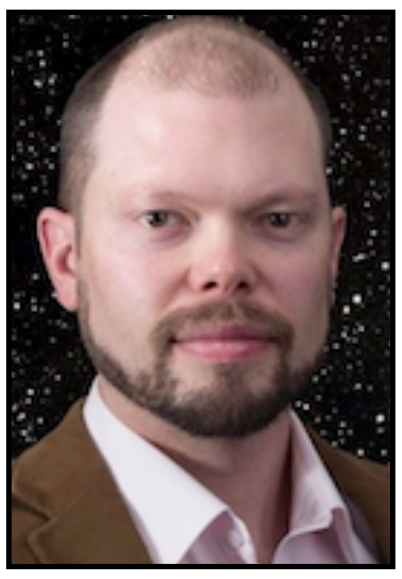

Dr Duane Hamacher is a Senior Australian Research Council Discovery Early Career Research Fellow at the Monash University Indigenous Centre in Melbourne. His research focuses on cultural astronomy, specialising in Indigenous Australia and Oceania. Duane leads the Indigenous Astronomy Group, is an Associate Editor of the *Journal of Astronomical History and Heritage*, and was elected to the Council of the International Society for Archaeoastronomy and Astronomy in Culture.

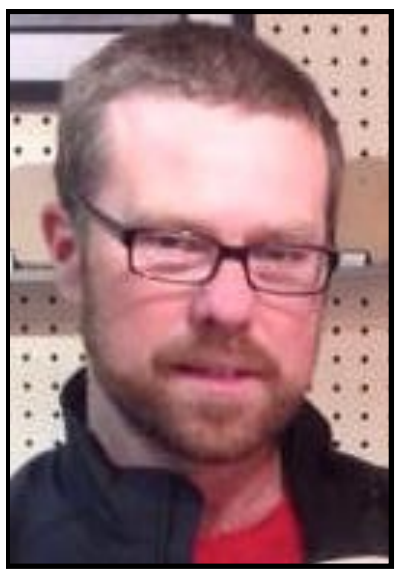

Mark Carter is an independent consultant zoologist in Alice Springs, Australia. He specialises in terrestrial vertebrate desert fauna and bioacoustics and has worked on diverse ecological research projects in the UK, North Africa, and the Australian arid-zone. His current research interests are tropical microbat taxa isolated in desert refugia in Australia and Oman, and bioacoustic bird survey methodologies. Mark is co-founder and Chair of *BirdLife Central Australia* and also owns and operates Birding and Wildlife.

# 11 Appendix

*Table A1: Stellar aspects (AR = acronychal rise; AS = acronychal set; HR = heliacal rise; HS = heliacal set; Mdawn = dawn meridional transit; Mdusk = dusk meridional transit) versus annual lifecycle stage (Mating/breeding; Birthing/hatching; whelping/fledging) for Dingo Mother (Achernar), Crow Mother (Altair), Black Cockatoo (Antares),and Red Back Spider (Arcturus). Bold colours denote peaks in lifecycle stages.*

| | | Star Aspect | | | | | | Dingo Mother - Achernar | | | Crow Mother - Altair | | | Black Cockatoo - Antares | | | Red Back Spider - Arcturus | |
|---|---|---|---|---|---|---|---|---|---|---|---|---|---|---|---|---|---|---|
| Month | Week | AR | AS | HR | HS | Mdawn | Mdusk | Mating | Birth | Whelping | Mating | Hatching | Fledging | Mating | Hatching | Fledging | Laying | Hatching |
| January | 1 | Crux | | | | | | | | | | | | | | | | |
| | 2 | | | | | | Pleiades | | | | | | | | | | | |
| | 3 | | | | | | | | | | | | | | | | | |
| | 4 | | | | | Crux | | | | | | | | | | | | |
| February | 1 | | | | | | | | | | | | | | | | | |
| | 2 | | | | | Arcturus | | | | | | | | | | | Mdawn | Mdawn |
| | 3 | | Canopus | Altair | | | | | | | HR | HR | HR | | | | | |
| | 4 | | | Vega | | | | | | | | | | | | | | |
| March | 1 | | | Delphinus | | | | | | | | | | | | | | |
| | 2 | | | Achernar | | Antares | Canopus | HR | HR | HR | | | | Mdawn | Mdawn | Mdawn | | |
| | 3 | | | | | | | | | | | | | | | | | |
| | 4 | | | | | | | | | | | | | | | | | |
| April | 1 | | | | | Vega | | | | | | | | | | | | |
| | 2 | | Arcturus | | | | | | | | | | | | | | AS | AS |
| | 3 | | | | Pleiades | | | | | | | | | | | | | |
| | 4 | | | | | Altair | | | | | Mdawn | Mdawn | Mdawn | | | | | |
| May | 1 | | | | | | | | | | | | | | | | | |
| | 2 | Arcturus | | | | Delphinus | | | | | | | | | | | AR | AR |
| | 3 | Antares | Crux | | | | | | | | | | | AR | AR | AR | | |
| | 4 | | Vega | Canopus | Achernar | | | HS | HS | HS | | | | | | | | |
| June | 1 | | | | | | | | | | | | | | | | | |
| | 2 | | Antares | | | | | | | | | | | AS | AS | AS | | |
| | 3 | | | Pleiades | | | | | | | | | | | | | | |
| | 4 | | | | | | | | | | | | | | | | | |
| July | 1 | | Altair | | | | Crux | | | | AS | AS | AS | | | | | |
| | 2 | | | | | | | | | | | | | | | | | |
| | 3 | | Delphinus | | | Achernar | Arcturus | Mdawn | Mdawn | Mdawn | | | | | | | Mdusk | Mdusk |
| | 4 | Altair | | | | | | | | | AR | AR | AR | | | | | |
| August | 1 | Vega/Delph | | | Canopus | | | | | | | | | | | | | |
| | 2 | | | | | | | | | | | | | | | | | |
| | 3 | Achernar | | | | | Antares | AR | AR | AR | | | | Mdusk | Mdusk | Mdusk | | |
| | 4 | | | Crux | | | | | | | | | | | | | | |
| September | 1 | | | | | Pleiades | | | | | | | | | | | | |
| | 2 | | | | | | | | | | | | | | | | | |
| | 3 | | | | | | Vega | | | | | | | | | | | |
| | 4 | | | | Arcturus | | | | | | | | | | | | HS | HS |
| October | 1 | | | | | | Altair | | | | Mdusk | Mdusk | Mdusk | | | | | |
| | 2 | | | | | | Delphinus | | | | | | | | | | | |
| | 3 | | | | | Canopus | | | | | | | | | | | | |
| | 4 | | | | Crux | | | | | | | | | | | | | |
| November | 1 | Canopus | | | Vega | | | | | | | | | | | | | |
| | 2 | Peiades | | | | | | | | | | | | | | | | |
| | 3 | | | | Antares | | | | | | | | | HS | HS | HS | | |
| | 4 | | | | | | | | | | | | | | | | | |
| December | 1 | | Pleiades | | Altair | | | | | | | | | | | | | |
| | 2 | | | Arcturus | Alt/Delph | | | | | | HS | HS | HS | | | | HR | HR |
| | 3 | | | Antares | | | Achernar | Mdusk | Mdusk | Mdusk | | | | HR | HR | HR | | |
| | 4 | | Achernar | | | | | AS | AS | AS | | | | | | | | |

*Table A1 (cont.): Stellar aspects (AR = acronychal rise; AS = acronychal set; HR = heliacal rise; HS = heliacal set; Mdawn = dawn meridional transit; Mdusk = dusk meridional transit) versus annual lifecycle stage (Mating/breeding; Birthing/hatching; whelping/ fledging, torpur) for Owlet Nightjar (Canopus), Wedge-Tailed Eagle (Crux), Crow Chicks (Delphinus), Thorny Devil (Pleiades), and Bustard (Vega ). Bold colours denote peaks in lifecycle stages.*

| | | Star Aspect | | | | | | Owlet Nightjar - Canopus | | | | Wedge Tail Eagle - Crux | | | Crow Chicks - Delphinus | | | Thorny Devil - Pleiades | | Bustard - Vega | | |
|---|---|---|---|---|---|---|---|---|---|---|---|---|---|---|---|---|---|---|---|---|---|---|
| Month | Week | AR | AS | HR | HS | Mdawn | Mdusk | Mating | Hatching | Fledging | Torpor | Mating | Hatching | Fledging | Mating | Hatching | Fledging | Laying | Hatching | Mating | Hatching | Fledging |
| January | 1 | Crux | | | | | | | | | | AR | AR | AR | | | | | | | | |
| | 2 | | | | | | Pleiades | | | | | | | | | | | Mdusk | Mdusk | | | |
| | 3 | | | | | | | | | | | | | | | | | | | | | |
| | 4 | | | | | Crux | | | | | | Mdawn | Mdawn | Mdawn | | | | | | | | |
| February | 1 | | | | | | | | | | | | | | | | | | | | | |
| | 2 | | | | | Arcturus | | | | | | | | | | | | | | | | |
| | 3 | | Canopus | Altair | | | | AS | AS | AS | AS | | | | | | | | | | | |
| | 4 | | | Vega | | | | | | | | | | | | | | | | HR | HR | HR |
| March | 1 | | | Delphinus | | | | | | | | | | | HR | HR | HR | | | | | |
| | 2 | | | Achernar | | Antares | Canopus | Mdusk | Mdusk | Mdusk | Mdusk | | | | | | | | | | | |
| | 3 | | | | | | | | | | | | | | | | | | | | | |
| | 4 | | | | | | | | | | | | | | | | | | | | | |
| April | 1 | | | | | Vega | | | | | | | | | | | | | | Mdawn | Mdawn | Mdawn |
| | 2 | | Arcturus | | | | | | | | | | | | | | | | | | | |
| | 3 | | | | Pleiades | | | | | | | | | | | | | HS | HS | | | |
| | 4 | | | | | Altair | | | | | | | | | | | | | | | | |
| May | 1 | | | | | | | | | | | | | | | | | | | | | |
| | 2 | Arcturus | | | | Delphinus | | | | | | | | | Mdawn | Mdawn | Mdawn | | | | | |
| | 3 | Antares | Crux | | | | | | | | | AS | AS | AS | | | | | | | | |
| | 4 | | Vega | Canopus | Achernar | | | HR | HR | HR | HR | | | | | | | | | AS | AS | AS |
| June | 1 | | | | | | | | | | | | | | | | | | | | | |
| | 2 | | Antares | | | | | | | | | | | | | | | | | | | |
| | 3 | | | Pleiades | | | | | | | | | | | | | | HR | HR | | | |
| | 4 | | | | | | | | | | | Mdusk | Mdusk | Mdusk | | | | | | | | |
| July | 1 | | Altair | | | | Crux | | | | | | | | | | | | | | | |
| | 2 | | | | | | | | | | | | | | | | | | | | | |
| | 3 | | Delphinus | | | Achernar | Arcturus | | | | | | | | AS | AS | AS | | | | | |
| | 4 | Altair | | | | | | | | | | | | | | | | | | | | |
| August | 1 | Vega/Delph | | | Canopus | | | HS | HS | HS | HS | | | | AR | AR | AR | | | AR | AR | AR |
| | 2 | | | | | | | | | | | | | | | | | | | | | |
| | 3 | Achernar | | | | | Antares | | | | | | | | | | | | | | | |
| | 4 | | | Crux | | | | | | | | HR | HR | HR | | | | | | | | |
| September | 1 | | | | | Pleiades | | | | | | | | | | | | Mdawn | Mdawn | | | |
| | 2 | | | | | | | | | | | | | | | | | | | | | |
| | 3 | | | | | | Vega | | | | | | | | | | | | | Mdusk | Mdusk | Mdusk |
| | 4 | | | | Arcturus | | | | | | | | | | | | | | | | | |
| October | 1 | | | | | | Altair | | | | | | | | | | | | | | | |
| | 2 | | | | | | Delphinus | | | | | | | | Mdusk | Mdusk | Mdusk | | | | | |
| | 3 | | | | | Canopus | | Mdawn | Mdawn | Mdawn | Mdawn | | | | | | | | | | | |
| | 4 | | | | Crux | | | | | | | HS | HS | HS | | | | | | | | |
| November | 1 | Canopus | | | Vega | | | AR | AR | AR | AR | | | | | | | | | HS | HS | HS |
| | 2 | Peiades | | | | | | | | | | | | | | | | AR | AR | | | |
| | 3 | | | | Antares | | | | | | | | | | | | | | | | | |
| | 4 | | | | | | | | | | | | | | | | | | | | | |
| December | 1 | | Pleiades | | Altair | | | | | | | | | | | | | AS | AS | | | |
| | 2 | | | Arcturus | Alt/Delph | | | | | | | | | | HS | HS | HS | | | | | |
| | 3 | | | Antares | | | Achernar | | | | | | | | | | | | | | | |
| | 4 | | Achernar | | | | | | | | | | | | | | | | | | | |